\documentclass[reqno]{amsart}
\usepackage{graphicx}

\newtheorem{theorem}{Theorem}

\newtheorem{remark}{Remark}

\newcommand{\be}{\begin{eqnarray}}
\newcommand{\ee}{\end{eqnarray}}
\newcommand{\bee}{\begin{eqnarray*}}
\newcommand{\eee}{\end{eqnarray*}}
\newcommand{\R}{{\mathbb R}}

\begin{document}

\title []{Stationary solutions of a fractional Laplacian with singular perturbation}

\author {Andrea Sacchetti}

\address {Department of Physics, Informatics and Mathematics, University of Modena e Reggio Emilia, Modena, Italy}

\email {andrea.sacchetti@unimore.it}

\date {\today}

\thanks {I'm very grateful to Diego Noja and Sandro Teta for useful discussions. \ This work is partially 
supported by Gruppo Nazionale per la Fisica Matematica (GNFM-INdAM)}

\begin {abstract} In this report we extend some ideas already developed by \cite {LRSRM,OCV,OV} to the case 
where the singular perturbation is given by a derivative of the Dirac's $\delta$.
\end {abstract}

\maketitle

\bigskip

{\it Ams classification (MSC 2010): 81Qxx}  

\bigskip

\section {Introduction} 

Since the seminal papers by Laskin (see \cite {La1} and the reference therein) fractional quantum mechanics has 
received and increasing interest from a theoretical point of view, and only recently some applications have been 
proposed in optics \cite {Lo1,ZLBZZX} and in the framework of nonlinear Schr\"odinger equations \cite {ANV,CFHMT,KSM}. \ In fact, most of the 
current studies on 
fractional quantum mechanics are mainly focused on the mathematical aspects; the difficulties come from the fact that 
the fractional Laplacian $(-\Delta )^{\alpha /2}$ is a nonlocal operator. \ Therefore, analysis of some simple toy models would be very useful; 
recently, one-dimensional fractional Laplacian perturbed by a one (or more) Dirac's delta has been discussed \cite 
{LRSRM,OCV,OV}. 

In this short report we show how the main ideas can be extended to the case of a more singular 
perturbation of the one-dimensional fractional Laplacian, e.g. the $n$-th  
derivative of the Dirac's delta with $n \ge 1$; the price we have of pay is to request that the power $\alpha$ of the fractional 
Laplacial must be greater than $2n+1$. \ Hence this method does not apply to the standard Laplacian (corresponding to 
$\alpha =2$) when $n$ is bigger or equal than $1$. This problem does not occur for the Dirac's delta, corresponding to $n=0$, because it is infinitesimally bounded with respect 
to the Laplaclian (see Theorem KLMN and Example 3 by \S X.2 \cite {RS2}). \ In fact, one could extend the result to $\alpha \le 2n +1$, even for $n \ge 1$, 
by means of a suitable renormalization procedure as done, for example, by \cite {AGHHE} in order to define the three-dimensional Laplacian perturbed 
by a Dirac's delta.

In Section 2 we give the main result, and in Section 3 we apply it to the case of $n=0$ and $n=1$. \ In fact, the result 
when $n=0$ coincides with the one already given by \cite {OCV,OV} up to a normalization pre-factor. 

\section {Main result}

Let us consider the one-dimensional fractional eigenvalue problem $H\psi = E\psi$, $\psi \in L^2 (\R , dx)$, 
\be
H \psi := \left ( -\Delta \right )^{\alpha /2} \psi + V_0 \delta^{(n)} \psi \, ; \label {F0}
\ee
where $\delta^{(n)}$ is the $n$-th derivative of the Dirac's delta with strength $V_0 \in \R$. \ The fractional Laplacian operator is defined for 
any $\alpha >0$ by means of the Fourier transform ${\mathcal F}$ of $\psi$, that is 
\be
{\mathcal F} \left [ \left ( - \Delta \right )^{\alpha /2} \psi \right ] (p) = |p|^{\alpha } \phi (p) , \ \phi = {\mathcal F} \psi \, , \label {F1}
\ee
and the $n$-th derivative of the Dirac's delta is defined as usual
\bee
\int_{\R} \delta^{(n)} f(x) dx = (-1)^n \frac {d^n f (0)}{dx^n} 
\eee
for any test function $f$ and its Fourier transform is given by $ {\mathcal F} \left ( \delta^{(n)} \right ) 
= (ip)^n$. \ Here
\bee
\phi (p) = \left [ {\mathcal F}\psi \right ] (p) = \int_{\R} e^{-i p x} \psi (x) dx 
\eee
and 
\bee 
\psi (x) = \left [ {\mathcal F}^{-1} \phi \right ] (x) = \frac {1}{2\pi} \int_{\R} e^{i p x} \phi (p) dp\, .
\eee

Laplacian fractional operators may have equivalent definitions provided that $\alpha \in (0,2)$ \cite {K}; 
furthermore, when $\alpha $ is restricted to such an interval 
then properties of regularity to the equation $(-\Delta )^{\alpha /2} u = f$ occurs. \ In these paper we don't 
restrict $\alpha$ to the interval $(0,2)$, indeed definition 
(\ref {F1}) makes sense even for $\alpha$ large enough provided that $\phi$ rapidly decreases when $p$ goes to infinity; in fact, we will assume $\alpha$ 
larger that $2n+1$.

If we consider the Fourier transform of both sides of the eigenvalue equation $H\psi = E\psi$ we have that it takes the form  
\bee
|p|^{\alpha } \phi + V_0 {\mathcal F}\left ( \delta^{(n)} \right ) \star {\mathcal F} (\psi )= E \phi 
\eee
that is
\be
|p|^{\alpha } \phi + i^n \frac {V_0}{2\pi} \int_{\R} (p-q)^n \phi (q)  dq = E \phi \, . \label {F0Bis}
\ee
In order to solve such an equation we set 
\be
K_h = - i^n (-1)^{n-h} \frac {V_0}{2\pi} \left ( \begin {array}{c} n \\ h \end {array} \right )  \int_{\R} q^{n-h} \phi (q) dq \, , \ h=0,1,2,\ldots ,\ n\, ; \label {F2}
\ee
hence it follows that (\ref {F0Bis}) has solution 
\be
\phi (p) = \sum_{h=0}^n K_h \frac { p^h }{|p|^{\alpha} -E} \, . \label {F3}
\ee
We should remark that $\phi \in L^2$ when $E<0$ and $\alpha > n+\frac {1}{2}$; furthermore the integrals in $K_h$ converge for any $h=0,\ldots ,n$ when 
$\alpha > 2n+1$. \ Therefore, by substituting (\ref {F3}) inside (\ref {F2}) we have that
\be
K_h = \sum_{k=0}^n a_{h,k} K_k \label {F3Bis}
\ee 
where (see \S \ref {IA})
\be
a_{h,k} &:=& a_{h,k}(E) = - i^n (-1)^{n-h} \frac {V_0}{2\pi} \left ( \begin {array}{c} n \\ h \end {array} \right ) 
\int_{\R} 
\frac {q^{n+k-h}}{|q|^{\alpha} + |E|} dq \label {IntA} \\ 
&=& - i^n (-1)^{n-h} \left ( \begin {array}{c} n \\ h \end {array} \right )
|E|^{\frac {n+k-h+1-\alpha}{\alpha}}\frac {V_0[1+(-1)^{n+k-h}] }{2\alpha \sin \left (\pi \frac {n+k-h+1}{\alpha}\right )}
\, . \nonumber
\ee
Then (\ref {F3Bis}) can be seen as a linear system $\sum_{k=0}^n \left [ a_{h,k} - \delta_h^k \right ] K_k =0$ which has no identically zero solution 
provided the associated matrix has determinant equal to zero. \ In fact, the eigenvalue equation is finally given by
\be
\mbox { det } \left [ a_{h,k}(E) - \delta_h^k \right ] =0 \, . \label {F4}
\ee
Once one has obtained the solutions $\hat E$ to equation (\ref {F4}) then the eigenvectors $\psi (x)$ are given by the inverse Fourier transform of (\ref {F3}), 
that is 
\be
\psi (x) &=&  
\sum_{h=0}^n \hat K_h \frac {1}{2\pi} \int_{\R}  \frac { p^h }{ |p|^{\alpha}+ |\hat E|} e^{i p x} dp \nonumber \\ 
&=& \sum_{h=0}^n \hat K_h \frac {1}{2\pi i^h} \frac {d^h}{dx^h}  F_\alpha (x) \label {IntB}
\ee
where
\bee
F_\alpha (x) := \int_{\R}  \frac { 1 }{|p|^{\alpha}+|\hat E|}  e^{i p x} dp
\eee
and where $\{ \hat K_h \}_{h=0}^n$ is a solution to the linear system $\sum_{k=0}^n \left [ a_{h,k} (\hat E) - \delta_h^k \right ] K_k =0$ under the normalization 
condition $\int_{\R} |\phi (p)|^2 d p =2\pi $ where $\phi (p) $ is given by (\ref {F3}), that is 
\bee
2\pi = \sum_{h,k=0}^n \overline {\hat K_h} \hat K_k M_{h+k,\alpha} (\hat E)
\eee
where we define
\be
  M_{h+k,\alpha} (E) := \int_{\R} \frac {p^{h+k}}{\left [  |p|^{\alpha } + | E | \right ]^2} d p \, .  \label {IntC}
\ee
For an explicit expression of $F_\alpha (x)$ and $M_{m,\alpha }(E)$ see, respectively, \S \ref {IB} and \S \ref {IC}.

We can collect all these results within the following statement.

\begin {theorem}
Let us consider the eigenvalue equation $H\psi = E\psi$, where $H$ is formally defined in $L^2 (\R , dx)$ by (\ref {F0}), with $\alpha > 2 n +1$. \ Then, the real and 
negative eigenvalues are solutions to equation (\ref {F4}), where $a_{h,k}$ are defined by (\ref {IntA}), with associated normalized eigenvectors 
(\ref {IntB}).
\end {theorem}

\section {Examples} 

\subsection {Dirac's $\delta$: $n=0$} 
\begin{center}
\begin{figure}
\includegraphics[height=8cm,width=8cm]{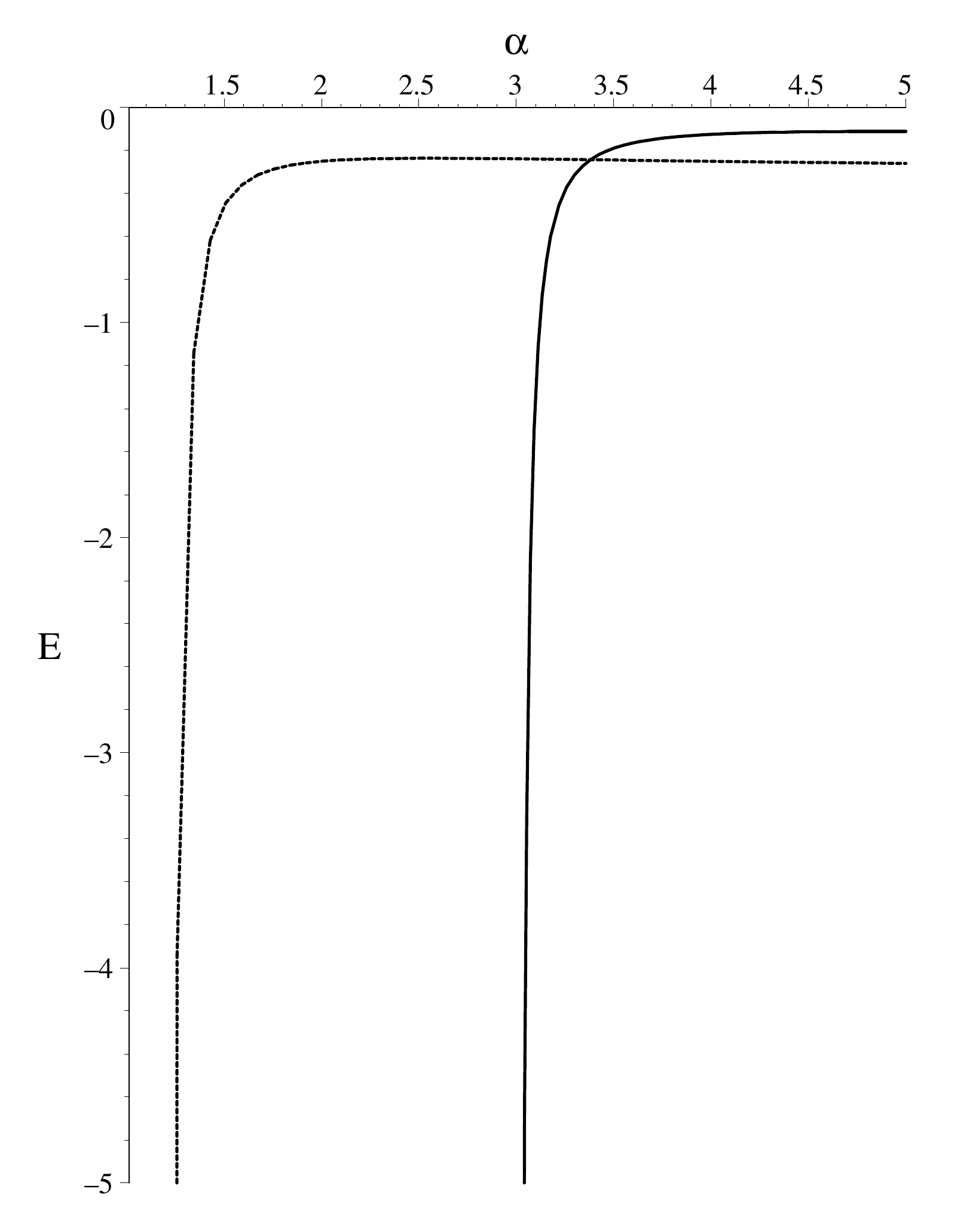}
\caption{\label {Figura1} Here we plot the eigenvalue $\hat E = \hat E (\alpha )$. \ Point line corresponds to the eigenvalue 
in the Dirac's delta case (\ref {Eq11}), for $\alpha >1$. \ Full line corresponds to the eigenvalue 
in the case (\ref {Eq12}) of the derivative of the Dirac's delta, for $\alpha >3$.}
\end{figure}
\end{center}

In such a case we assume that $\alpha > 1$ and 
the eigenvalues $\hat E$ are the real and negative 
solutions to the equation
\bee
a_{0,0}-1=0 \ \mbox { where } \ a_{0,0} = - \frac {V_0}{\alpha \sin \left ( \frac {\pi}{\alpha }
\right ) } |E|^{\frac {1-\alpha}{\alpha}} \, .
\eee
This equation has solution when $V_0 <0$ and it is given by (see point line in Figure \ref {Figura1})
\be
\hat E 
= - \left [ - \frac {V_0}{\alpha \sin  \left ( \frac {\pi}{\alpha } \right ) }\right ]^{\frac {\alpha}{\alpha-1}} \ . \label {Eq11}
\ee

\begin{center}
\begin{figure}
\includegraphics[height=8cm,width=8cm]{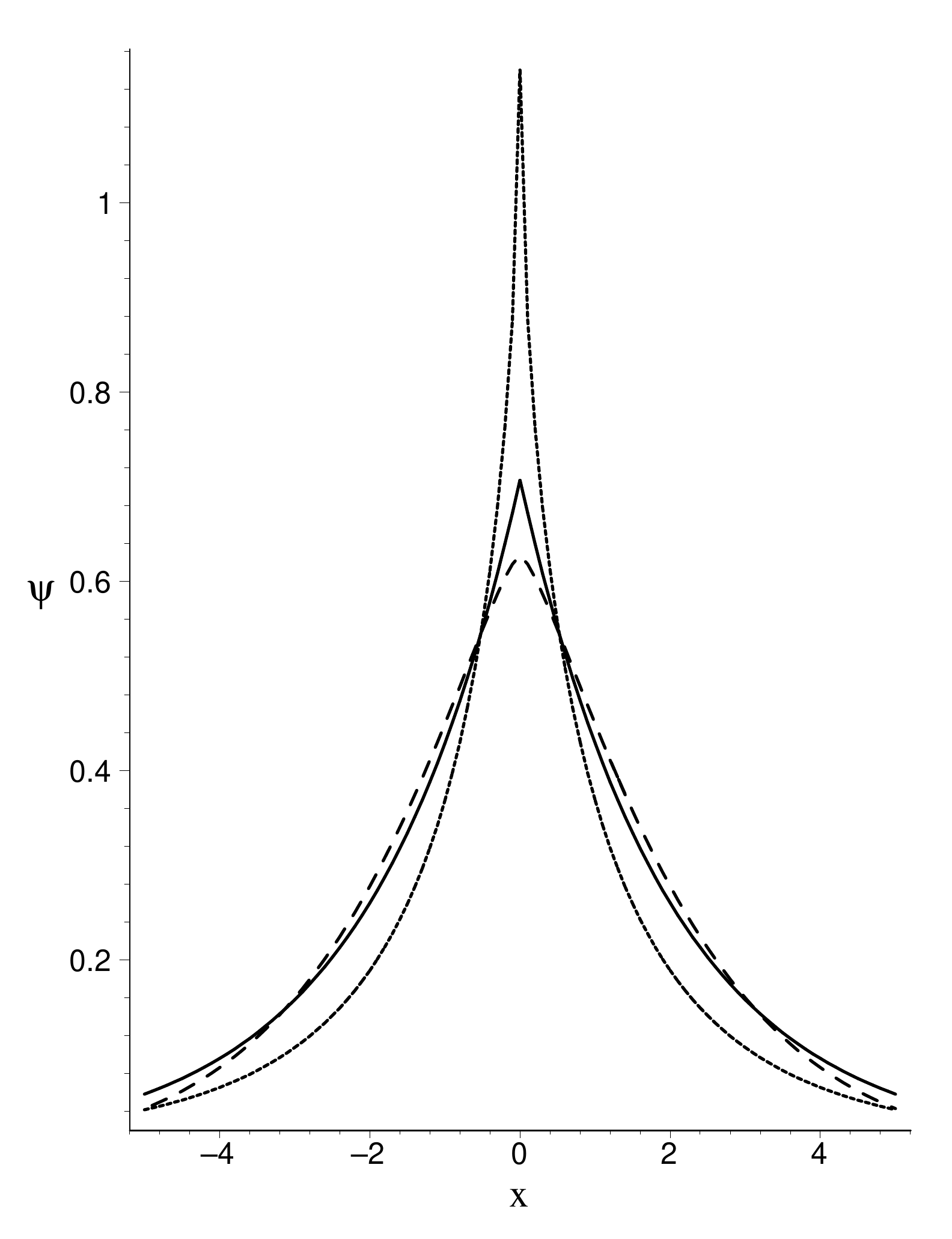}
\caption{\label {Figura2} Here we plot the normalized eigenvectors (\ref {F17}) for different values of $\alpha$ 
(point line corresponds to $\alpha =1.5$, full line corresponds to $\alpha =2$ and broken line corresponds to 
$\alpha =2.5$).}
\end{figure}
\end{center}

Concerning the normalized eigenvector we have that $\phi (p) = \hat K_0 \frac {1}{|\hat E | + |p|^{\alpha }}$ 
where $\hat K_0$ is such that $|\hat K_0 |^2 M_{0,\alpha} (\hat E) =2 \pi $; that is 
\bee
\hat K_0 = \left [ \frac {2\pi}{M_{0,\alpha} (\hat E)} \right ]^{\frac 12} 
= \left [ - V_0 \frac {\alpha}{\alpha -1} |\hat E| \right ]^{\frac 12 }
\, ,
\eee
by \S \ref {IC}. \ Hence, by \S \ref {IB}, the normalized eigenvector is given by (see Figure \ref {Figura2})
\be
\psi (x) 
&=& \hat K_0 \frac {1}{2\pi} F_\alpha (x) \nonumber \\
&=& \left [  \frac {- {V_0} \alpha }{(\alpha -1)|\hat E|} \right ]^{\frac 12}  \frac {1}{|x|} 
H_{2,3}^{2,1} \left [ |\hat E| \, |x|^{\alpha } 
\left | 
\begin {array}{lll}
 (1,1),&\left (1,\frac {\alpha}{2} \right )& \\ (1,\alpha ),& (1,1),& \left (1,\frac {\alpha}{2} \right )
\end {array}
\right. 
\right ] \, . \label {F17}
\ee

\begin {remark}
In fact, this result has been already given by \cite {OCV}, eqns. (35) and (36), with a slightly different normalization coefficient, which coincides with the one 
given by (\ref {F17}) when $\alpha =2$.
\end {remark}

\subsection {Derivative of the Dirac's $\delta$: $n=1$}

In such a case we assume that $\alpha >3$. \ One has that $a_{0,0}=a_{1,1}=0$ and that 
\bee
a_{0,1} = i \frac {V_0}{\alpha \sin \left ( \frac {3\pi}{\alpha} \right )} |E|^{(3-\alpha)/\alpha} \ \mbox { and } \ 
a_{1,0} = -i \frac {V_0}{\alpha \sin \left ( \frac {\pi}{\alpha} \right )} |E|^{(1-\alpha)/\alpha}
\eee
Hence (\ref {F3Bis}) takes the form
\bee
\left \{
\begin {array}{lcl}
 K_0 &=& a_{0,1} K_1 \\
 K_1 &=& a_{1,0} K_0
\end {array}
\right.
\eee
and the real and negative eigenvalues $\hat E$ are the solutions to the equation $a_{0,1}a_{1,0}=1$, that is (see full line in Figure \ref {Figura1})
\be
\hat E 
= - \left [ \frac {|V_0|} {\alpha \left [ \sin \left ( \frac {3\pi}{\alpha} \right ) 
\sin \left ( \frac {\pi}{\alpha} \right ) \right ]^{1/2}} \right ]^{\alpha/(\alpha-2)}\label {Eq12}
\ee
We may remark that the eigenvalue equation has solutions $\hat E := \hat E(V_0)$ for any $V_0 \in \R$, and in particular 
that $\hat E(-V_0 ) =\hat E(V_0)$.

\begin{center}
\begin{figure}
\includegraphics[height=8cm,width=8cm]{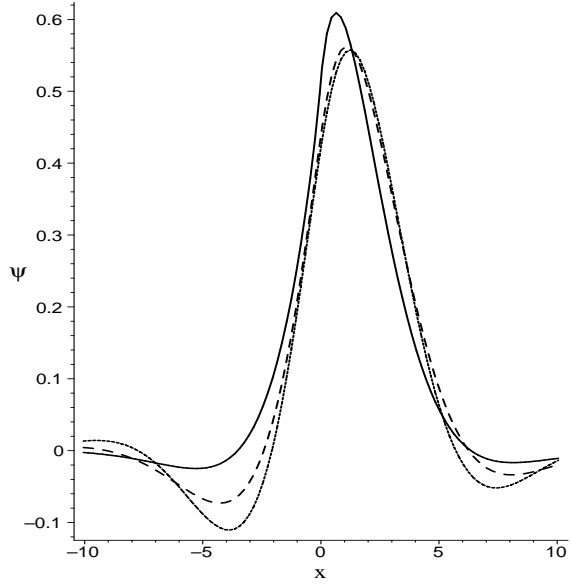}
\caption{\label {Figura3} Here we plot the normalized eigenvectors (\ref {Eq13}) for different values of $\alpha$ 
(full line corresponds to $\alpha =3.5$, broken line corresponds to $\alpha =4$ and point line corresponds to 
$\alpha =5.5$).}
\end{figure}
\end{center}

Concerning the normalized eigenvector we have that $\phi (p) = \frac {\hat K_0 + p \hat K_1}{|p|^\alpha + |\hat E|}$ and where $\hat K_0 =c$ and $\hat K_1 = a_{1,0} (\hat E) c$ where $c$ 
is a normalization constant given by 
\bee
c = \sqrt {2\pi} \left [ M_{0,\alpha } (\hat E) + |a_{1,0}(\hat E) |^2 M_{2,\alpha }(\hat E) \right ]^{-1/2}
\eee
Hence the eigenvector $\psi (x)$ is given by (see Figure \ref {Figura3})
\be
\psi (x) = \frac {c}{2\pi} F_\alpha^0 (x) + \frac {c}{2\pi i} a_{1,0}(\hat E) F_\alpha^1 (x) \label {Eq13}
\ee
where $F_\alpha^0 (x)$ is an even function defined by $F_\alpha (x)$ for $x\ge 0$, and where $F_\alpha^1 (x) $ is an odd function 
defined as $F_\alpha^1 (x) = \frac {dF_\alpha }{dx}$ for $x>0$, i.e. (see \S \ref {IB}):
\bee
F_\alpha^1 (x) := \frac {2\pi }{|E| x^2} H^{2,2}_{3,4} \left [ |E| x^{\alpha} \left | 
\begin {array}{cccc} 
(1,\alpha ),&(1,1),& (1,\alpha/2), &  \\ 
(1,\alpha),& (1,1) , &(1,\alpha/2),&(2,\alpha ) 
\end{array}
\right.
\right ]\, ,\ x >0\, . 
\eee

\appendix

\section {Some integrals}

\subsection {Integral (\ref {IntA})} \label {IA}

In previous expressions we consider the integral
\bee
J_{m,\alpha } (E) := \int_{\R} \frac {q^m}{|q|^{\alpha }+|E|} dq 
= [1+(-1)^m] |E|^{\frac {m+1-\alpha}{\alpha}} \hat J_{m,\alpha} 
\eee
where
\bee
\hat J_{m,\alpha} := \int_0^{+\infty} \frac {w^m}{w^{\alpha }+1} dw \, . 
\eee
Recalling that
\bee
I_{( z , y )} (p) := \int \frac {p^y}{p^z+1 } dp = \frac {p^{1+y}}{y+1} F  
\left ( \left [1, \frac yz + \frac 1z \right ], \left [1+ \frac yz + \frac 1z \right ] , -p^z \right )
\eee
where $F \left ( \left [ a,b \right ] , \left [ c \right ] ; z \right )$ is the hypergeometric 
function. \ Then
\bee 
\hat J_{m,\alpha } = \lim_{p\to +\infty} 
\frac {p^{1+m}}{m+1} F 
\left ( \left [1, \frac {m+1}{\alpha} \right ], \left [1+ \frac {m+1}{\alpha} \right ] 
, -p^{\alpha}  \right ) 
\eee
Now, recalling formula (15.3.7) by \cite {AS} which holds true when $|\mbox {arg} (-z)| < \pi$, it follows that 
\bee 
F \left ( \left [ a,b \right ] , \left [ c \right ] ; z \right ) 
&=& \frac {\Gamma (c) \Gamma (b-a)}{\Gamma (b)\Gamma (c-a)} (-z)^{-a} 
F \left ( \left [ a, 1-c+a \right ] , \left [ 1-b+a \right ] ; z^{-1} \right ) + \\ 
&& \ \ + \frac {\Gamma (c) \Gamma (a-b)}{\Gamma (a)\Gamma (c-b)} (-z)^{-b} 
F \left ( \left [ b, 1-c+b \right ] , \left [ 1-a+b \right ] ; z^{-1} \right ) \, , 
\eee
hence 
\bee
&& \frac {p^{1+m}}{m+1} F 
\left ( \left [1, \frac {m+1}{\alpha} \right ], \left [1+ \frac {m+1}{\alpha} \right ] 
, -p^{\alpha}  \right ) = \\ 
&& \ \ =   \frac {p^{1+m}}{m+1} \frac {\Gamma \left (\frac {m+1}{\alpha} +1 \right ) 
\Gamma \left (\frac {m+1}{\alpha} -1 \right )}{\Gamma^2 \left (\frac {m+1}{\alpha} \right )} 
p^{-\alpha} 
F \left ( \left [ 1, 1-\frac {m+1}{\alpha} \right ] , \left [  2-\frac {m+1}{\alpha} \right ] ; 
(-p)^{-\alpha} \right ) + \\
&& \ \ +  \frac {p^{1+m}}{m+1} \frac {\Gamma \left (\frac {m+1}{\alpha} +1 \right ) 
\Gamma \left (1-\frac {m+1}{\alpha}  \right )}{\Gamma^2 \left (1 \right )} 
p^{-(m+1)} 
F \left ( \left [ \frac {m+1}{\alpha}, 0 \right ] , \left [  \frac {m+1}{\alpha} \right ] ; 
(-p)^{-\alpha} \right ) \, . 
\eee
Therefore
\bee
\hat J_{m,\alpha} = \frac {1}{m+1} \frac {\Gamma \left (\frac {m+1}{\alpha} +1 \right ) 
\Gamma \left (1-\frac {m+1}{\alpha}  \right )}{\Gamma^2 \left (1 \right )} = 
\frac {\pi }{\alpha \sin \left (\pi \frac {m+1}{\alpha}\right )}
\eee
from which follows that 
\bee 
J_{m,\alpha } (E) =  |E|^{\frac {m+1-\alpha}{\alpha}}\frac {[1+(-1)^m]\pi }{\alpha \sin \left (\pi \frac {m+1}{\alpha}\right )}
\eee
provided that $\alpha > m+1$.

\subsection {Integral (\ref {IntC})} \label {IC}
Now we consider the integral 
\bee
M_{m,\alpha } (E) := \int_{\R} \frac {q^m}{\left [ |q|^{\alpha }+ |E|\right ]^2} dq 
= [1+(-1)^m] |E|^{\frac {m+1-2\alpha}{\alpha}} \hat M_{m,\alpha }
\eee
where
\bee
\hat M_{m,\alpha } := 
\int_0^{+\infty} \frac {w^m}{\left [ w^{\alpha }+1\right ]^2} dw \, . 
\eee
Observing that
\bee
\int \frac {p^y}{(p^z+1)^2 } dp = + \frac 1z \frac {p^{y+1}}{p^z+1} - \frac {y+1-z}{z} \int \frac {p^{y}}{p^z+1} dp 
\eee
then 
\bee
\hat M_{m,\alpha} = - \frac {m+1-\alpha}{\alpha} \hat J_{m,\alpha} = -\frac {m+1-\alpha}{\alpha} 
\frac {\pi}{\alpha \sin \left ( \pi \frac {m+1}{\alpha }\right ) }
\eee
under the conditions $\alpha > m+1$.

\subsection {Integral (\ref {IntB})} \label {IB}
Now we consider the integral 
\bee
F_\alpha (x):= \int_{\R}  \frac { 1 }{ |p|^{\alpha}+|E|} e^{i p x} dp 
\eee
This function is an even parity function, i.e. $F_\alpha (-x)=F_\alpha (x)$. \ By (B12) by \cite {OCV} it follows that
\bee
F_\alpha (x) = \frac {2\pi}{|E| \, |x|} 
H_{2,3}^{2,1} \left [ |E| \, |x|^{\alpha } 
\left | 
\begin {array}{lll}
 (1,1),&(1,\alpha /2 )& \\ (1,\alpha ),& (1,1),& (1,\alpha /2 ) 
\end {array}
\right. 
\right ]
\eee
where $H$ denotes the Fox's special function. \ Here we recall that the Fox's special functions are defined as follows \cite {MSH}
\bee
H^{m,n}_{p,q} \left [ z
\left | 
\begin {array}{lll}
 (a_1,A_1),&\cdots ,& (a_p,A_p ) \\ (b_1,B_1 ),& \cdots ,& (b_q,B_q ) 
\end {array}
\right. 
\right ] := \left [ {\mathcal M}^{-1} \Theta \right ] (z)
\eee
where ${\mathcal M}^{-1}$ denotes the inverse Mellin's transform and where 
\bee
\Theta (s) = \frac {\Pi_{j=1}^m \Gamma (b_j + B_j s) \Pi_{\ell =1}^n \Gamma (1-a_\ell -A_\ell s )}
{\Pi_{j=m+1}^q \Gamma (1-b_j - B_j s) \Pi_{\ell =n+1}^n \Gamma (a_\ell +A_\ell s )}
\eee
Finally, we recall the following formula concerning the derivative of the $H$-function (see formula (1.83) by \cite {MSH}):
\bee
&& 
\frac {d}{dx} \left \{ x^s H^{m,n}_{p,q} \left [ z x^h \left | 
\begin {array}{ccc} 
(a_1,A_1),& \cdots , &(a_p,A_p) \\ 
(b_1,B_1),& \cdots , &(b_q,B_q) 
\end{array}
\right.
\right ]
\right \} = \\
&& \ \ = 
x^{s-1} H^{m,n+1}_{p+1,q+1} \left [ z x^h \left | 
\begin {array}{cccc} 
(-s,h),&(a_1,A_1),& \cdots , &(a_p,A_p) \\ 
(b_1,B_1),& \cdots , &(b_q,B_q),&(-s+1,h) 
\end{array}
\right.
\right ]
\eee
provided that $h>0$.


\begin{thebibliography}{0} 

\bibitem {AS}  M. Abramowitz, and I.A. Stegun, {\it Handbook of mathematical functions with formulas, 
graphs, and mathematical tables}, Wiley (1972).

\bibitem {ANV} R. Adami, D. Noja, and N. Visciglia, {\it Constrained energy minimization and ground states for NLS with point defects}, 
Disc. Cont. Dyn. Syst. B {\bf 18}, 1155-1188 (2013).

\bibitem {AGHHE} S. Albeverio, F. Gestezy, R. Hoegh-Kronn, H. Holden, e P. Exner, {\it Solvable models in Quantum
Mechanics}, AMS Chelsea Publ. (second edition 2004).

\bibitem {CFHMT} Y. Cho, M.M. Fall, H. Hajaiej, P.A. Markowich, S. Trabelsi, {Orbital stability of standing 
waves of a class of fractional Schr\"odinger equations with a general Hartree-type integrand}, Anal. Appl. {\bf 15}, 699 (2017).

\bibitem {KSM} C. Klein, C. Sparber, and P. Markowich, {\it Numerical study of fractional nonlinear Schr\"odinger 
equations}, Proc. R. Soc. A {\bf 470}, 20140364 (2017).
 
\bibitem {K} M. Kwa\'snicki, {\it Ten equivalent definitions of the fractional Laplace operator}, 
Fractional Calculus and Applied Analysis {\bf 20}, 7-51 (2017).

\bibitem {La1} N. Laskin, {\it Principles of fractional quantum mechanics}, in {\it Fractional Dynamics: Recent Advances} edited by  
J. Klafter,‎ S.C. Lim,‎ and R. Metzler, World Scientific Publishing Company (2011).

\bibitem {LRSRM} E.K. Lenzi, H.V. Ribeiro, M.A.F. dos Santos, R. Rossato,
and R.S. Mendes1, {\it Time dependent solutions for a fractional Schr\"odinger equation with delta potentials}, 
J. Math. Phys. {\bf 54}, 082107 (2013).

\bibitem {Lo1} S. Longhi, {\it Fractional Schr\"odinger equation in optics}, Opt. Lett. {\bf 40}, 1117-1120 (2015)

\bibitem {MSH} A.M. Mathai, R.K. Saxena, and H.J. Haubold, {\it The $H$-Function: theory and applications}, 
Spinger Verlag (2010)

\bibitem {OCV} E.C. de Oliveira, F.S. Costa, and J. Vaz, {\it The fractional Schr\"odinger equation for 
delta potentials}, J. Math. Phys. {\bf 51}, 123517 (2010).

\bibitem {OV} E.C. de Oliveira, and J. Vaz, {\it Tunneling in fractional quantum mechanics}, J. Phys. A: Math. Theor. {\bf 44}, 185303 (2011).

\bibitem {RS2} M. Reed, and B. Simon, {\it Methods of Modern Mathematical Physics II: Fourier analysis, self-adjointness}, Academic Press (1975).

\bibitem {ZLBZZX} Y. Zhang, X. Liu, M.R. Beli\'c, W. Zhong, Y. Zhang, and M. Xiao, {\it Propagation dynamics 
of a light beam in a fractional Schr\"odinger equation}, Phys. Rev. Lett. {\bf 115}, 180403 (2015).

\end{thebibliography}
\end {document}